\documentstyle[12pt]{article}
\begin{document}

\date{}

\title{{\bf Counterterms and dual holographic anomalies in CS gravity }}
\author{M\'aximo Ba\~nados$^{a}$, Rodrigo Olea$^{a}$ and Stefan Theisen$^{b}$ \medskip \medskip \\
\small \it $^{a}$Departamento de F\'{\i}sica, P. Universidad Cat\'{o}lica de Chile, Casilla 306, Santiago 22, Chile\\
\small \it $^{b}$Max-Planck-Institut f\"{u}r Gravitationphysik, Albert-Einstein-Institut, 14476 Golm, Germany }

\maketitle

\begin{abstract}
The holographic Weyl anomaly associated to Chern-Simons gravity in
$2n+1$ dimensions is proportional to the Euler term in $2n$
dimensions, with no contributions from the Weyl tensor. We compute
the holographic energy-momentum tensor associated to Chern-Simons
gravity directly from the action, in an arbitrary odd-dimensional
spacetime. We show, in particular, that the counterterms rendering
the action finite contain only terms of the Lovelock type.

\end{abstract}

The AdS/CFT correspondence\cite{Maldacena,GKP,Witten} provides a
prescription to compute vacuum expectation values of CFT operators
in terms of dual classical fields on AdS. This prescription has
been checked successfully in many examples.  In particular, the
CFT energy momentum tensor $T^{ij}$ is dual to the spacetime
metric governed by Einstein's equations with a negative cosmological
constant.  One can use the correspondence to compute the CFT Weyl
anomaly. This calculation was indicated in \cite{Witten}, and
carried on in detail in \cite{HS} with the expected result
\begin{equation}\label{A1}
\langle T^\mu_\mu\rangle= {\cal A} =  {N^2 \over 32\pi^2} \,
\left( R^{ij}R_{ij} - {1 \over 3} R^2 \right),
\end{equation}
for ${\cal N}=4$ $SU(N)$ SYM theory in four dimensions which is dual to type IIB string theory on
$AdS_5\times S^5$.

For a CFT (in $d=4$) with $n_s$ real scalars, $n_{f}$ Dirac spinors and $n_v$ vectors
the anomaly is \cite{Duff}
\begin{equation}\label{Agen}
{\cal A} = c\, C^2 -a \,E_4
\end{equation}
with \begin{eqnarray}
 a &=& {1 \over 360(4\pi)^2}\left(n_s + 11 n_f + 62 n_v \right)\\
 c &=& {1 \over 120(4\pi)^2}\left(n_s + 6n_f + 12 n_v \right),
\end{eqnarray}
and the curvature invariants $E_4$ and $C^2$ are
\begin{equation}\label{} E_4 = R^{ijkl}R_{ijkl}-4R^{ij}R_{ij}
+ R^2, \ \ \ \ \  C^2 = C^{ijkl}C_{ijkl}.
\end{equation}
Of course, (\ref{Agen}) reduces to (\ref{A1}) for the ${\cal N}=4$ multiplet with $n_s=6$, $n_f=2$ and $n_v=1$ \cite{HS}.

In higher (even) dimensions, the anomaly is always characterized by the coefficient
multiplying the Euler density and an increasing number of coefficients $c_i$
\footnote{$i=1$ for $d=4$, $i=1,2,3$ for $d=6$ \cite{BPB}, $i=1,\dots,12$
for $d=8$ \cite{BE}, while the number of Weyl-invariants in general dimension is still unknown.}
multiplying curvature invariants which transform homogeneously under Weyl rescaling of the metric.
This classification of anomalies has been identified in \cite{DS} where they were called type A and type B.

One general feature of any four-dimensional CFT whose dual gravity theory is
the Einstein-Hilbert action with cosmological constant is that
the two anomaly coefficients $a$ and $c$ are equal \cite{HS}. For a generic CFT
this is, however, not the case. Specific examples are the theories constructed
in \cite{ASTY}. This is reflected in the dual gravity theory by the presence of
higher derivative terms. For the theories of \cite{ASTY} they arise
from similar terms on the world-volume of D7 branes which are needed for their
construction.

Consider, for example, the following action in five dimensions
\begin{equation}\label{I5}
I = {1\over 2\kappa_5^2}\int d^{5}x\sqrt{-G}\left(\hat R - 2\Lambda +
\alpha \hat R^2 + \beta \hat R^{\mu\nu} \hat R_{\mu\nu} + \gamma
R^{\mu\nu\lambda\rho}  R_{\mu\nu\lambda\rho}  \right)
\end{equation}
Hatted objects refer to spacetime tensors. The Weyl anomaly,
associated to the CFT dual to (\ref{I5}) was calculated in
\cite{NojiriOdintsov,BlauNarainGava,ST} with the result,
\begin{equation}\label{Agen} {\cal A} ={R_{AdS_5}^2\over 16\kappa_5^2}
\left\lbrace(1 - 40\alpha - 8\beta + 4\gamma)C^2 -
(1 - 40\alpha - 8\beta - 4\gamma)E_4\right\rbrace.
\end{equation}
Using the AdS/CFT relations between the string theory and the gauge theory parameters
one finds for the theories constructed in \cite{ASTY}
that the coefficients in front of the higher derivative terms
are ${\cal O}(1/N)$ and consequently $a-c\sim N$ (with $a,c\sim N^2$).
However, for realistic CFT's such as (S)QCD inside its conformal window
one has $a-c\sim N^2$. So far no critical string theory dual for such theories
has been found (for attempts within the context of non-critical strings, see
\cite{KM,BCCKP}).

The higher derivative gravity theories we will consider in this note
cannot be considered as duals to realistic conformal field theories as they
are necessarily non-unitary \cite{BST}. But they exhibit peculiar features as far as the
anomalies are concerned since all type B anomalies vanish.

For the five-dimensional theory (arbitrary odd dimensions will be considered below),
Chern-Simons gravity arises at the point of enhanced gauge symmetry of
the action (\ref{I5}). This corresponds to the choice \footnote{%
The AdS radius $l$ has been set to unity.}
\begin{equation}\label{choice}
\Lambda =-3,\beta =-4\alpha=-4\gamma=-1,
\end{equation}
for which $I$ reduces to the five dimensional Chern-Simons form
\cite{Chamseddine}
\begin{equation}\label{CSA}
I =\int \mbox{Tr} \left(AdAdA + {3 \over 2} dA A^3 + {3 \over 5}
A^5  \right)
\end{equation} for the group $SO(4,2)$.

The particular choice (\ref{choice}) has several consequences
which were analyzed in detail in \cite{BST}. First, the
coefficient $c$ of the anomaly vanishes identically, as
it can be readily checked by inserting (\ref{choice}) in
(\ref{Agen}). The anomaly is then given by the the Euler term
alone. Since Chern-Simons gravities exist in all odd dimensions
\cite{Chamseddine,BTrZ,TrZ}, one may wonder whether the anomaly
associated to all Chern-Simons theories is the pure Euler term in
the corresponding dimension. This turns out to be true, as
indicated in \cite{BST} by an argument based on the equations of
motion.

The goal of this note is to prove this statement by
computing, directly from the action, the Chern-Simons-AdS
holographic energy momentum tensor and its corresponding anomaly
in any odd dimension $2n+1$.

We start by explaining the Hamiltonian method to compute the
holographic energy momentum tensor (see
\cite{Muck,KS,Verlinde2-deBoer} for other Hamiltonian approaches).
In the ADM parametrization using $r$ as ``time" in $D=d+1$
dimensions,
\begin{equation}
ds^{2} = N^2\, dr^{2} + h_{ij}(r,x)(dx^{i}+N^idr)(dx^{j}+N^j dr).
\end{equation}
the gravitational action can be cast in the Hamiltonian form
\footnote{ We choose to start with the Hamiltonian action for
convenience, but one may well start with the Lagrangian action. In
this context, it would be interesting to explore how the conformal
anomaly appears in the regularization scheme for CS-AdS gravity
action proposed in \cite{MOTZCS}. },
\begin{equation}\label{Ih}
I_0 = \int dr\int d^{d}x\left (\pi^{ij} h'_{ij} + N {\cal H} + N^i
{\cal H}_i \right).
\end{equation}
This form of the action is universal and follows by geometrical considerations. The explicit formulae for the
constraints vary depending on the particular theory. But for any invariant
theory of gravity, the action can
always be cast into the form (\ref{Ih}). The action $I_0$ needs boundary terms and counterterms to be well defined.
For the problem at hand, these can be computed as follows.

The on-shell variation of $I_0$ (assuming that the bulk fields satisfy the equations of motion) is
 \begin{equation}
\delta I_0   =   \int_{r=\varepsilon} d^dx\,  \pi^{ij}\, \delta
h_{ij}  \label{1}.
\end{equation}
Now, we make a definite choice for the lapse $N$ and shift $N^i$ functions at infinity, and assume the FG form
for the asymptotic metric
\begin{eqnarray}
ds^{2}= {dr^{2} \over 4r^2}  + \frac{1}{r}\, g_{ij}(x^i,r) dx^{i}dx^{j}
\label{ds2radial}
\end{eqnarray}%
which corresponds to $N={1\over 2r}$, $N^i=0$ and $h_{ij}={1\over
r}g_{ij}$. We shall not need to solve the asymptotic equations,
nor assume a particular expansion\footnote{For standard gravity
\cite{FG}, $g_{ij} \simeq g_{(0)ij} + r\,g_{(1)ij} +
r^2\,g_{(2)ij} \cdots$ . The coefficient $g_{(1)}$ is universal
and locally related to $g_{(0)}$. The energy momentum tensor
depends \cite{Myers,dHSS} on $g_{(n)}$ which is non-locally
related to $g_{(0)}$.} for $g_{ij}(r,x^i)$. We only assume that
the limit $r\rightarrow 0$,
\begin{equation}
g_{ij}(x^i,r) \ \longrightarrow \  g_{(0)ij}(x^i)
\end{equation}
exists. Under these conditions we will show that the variation (\ref{1}) can be rewritten in the form,
\begin{equation}
\delta I_0 =\int_{r=\varepsilon} d^dx\,\left({1\over 2}\sqrt{g_{(0)}}\ T^{ij}\,\delta g_{(0)ij}+\delta B \right) \label{2}
\end{equation}
where $T^{ij}$ is {\it finite}, and $B$ is a (divergent) local functional of $g_{ij}$.

From (\ref{2}) we conclude that the correct gravitational action
is obtained by passing the term $B$ to the left hand side. We
define the renormalized action
\begin{equation}\label{I}
I \equiv I_0 - \int d^dx\ B.
\end{equation}
From (\ref{2}) we see that its variation with respect to
$g_{(0)ij}$ is well defined and finite. Our goal is now to compute
the counterterm $B$, and the coefficient $T^{ij}$ which becomes
the holographic energy momentum tensor.

This procedure was carried out in \cite{BST} for Einstein gravity, and five-dimensional Chern-Simons gravity.
We shall now extend these results for Chern-Simons gravity in any odd dimension $D=2n+1$. In particular, we shall
compute the explicit formula for $B$, which turns out to be a Lovelock action in $2n$
dimensions\footnote{A note of caution is in order here. As shown in \cite{dHSS,Bianchi-}, when matter fields
are present, the counterterm action contributes to the finite piece in a non-trivial way. Here, we restrict
the discussion to the matter free action, and leave for the future a more general analysis.}.

It is worth mentioning here that the standard procedure to find the 1-point function (see \cite{Skenderis} for a review),
by solving the asymptotic equations, inverting the series, and varying with respect to the regularized metric
becomes unfeasible in Chern-Simons gravity due to the higher powers in the curvature tensor and the
resulting complicated
equations of motion. The method, described above, for finding the variation of the action becomes extremely powerful
if one deals with complicated actions of gravity.

Let us apply this procedure to Chern-Simons gravity in arbitrary dimensions.
Chern-Simons gravities are particular cases of Lovelock gravities. The Lovelock action is
\cite{Lovelock},
\begin{equation}
I=\sum\limits_{2p<D}\alpha _{p}I_{(p)}  \label{Ilovelock}
\end{equation}%
where the terms $I_{(p)}$%
\begin{equation}
I_{(p)}=\frac{1}{2p!}\int dr\,d^{d}x\sqrt{-G}\delta _{\left[ \mu
_{1}...\mu _{2p}\right] }^{\left[ \mu _{1}...\mu _{2p}\right]
}\hat{R}_{\mu _{1}\mu _{2}}^{\mu _{1}\mu _{2}}...\hat{R}_{\mu
_{2p-1}\mu _{2p}}^{\mu _{2p-1}\mu _{2p}} \label{Ip}
\end{equation}%
represent the dimensional continuation of the Euler densities of
the lower dimensions.

The Hamiltonian structure of this action was studied in \cite{TZ}. For this theory, the ``velocities"
$h'_{ij}$ cannot be inverted as functions of the momenta. But the relation  $\pi^{ij}(h'_{kl})$ does exists \cite{TZ},
\begin{equation}
\pi _{j}^{i}=\frac{1}{4}\sum\limits_{p\geq 0}\alpha
_{p}\sum\limits_{s=0}^{p-1}C_{s(p)}\left( \pi _{s(p)}\right)
_{j}^{i} \label{PIij}
\end{equation}%
where
\begin{equation}
\left( \pi _{s(p)}\right) _{j}^{i}=\sqrt{-h}\delta _{\left[ jj_{1}...j_{2s}...j_{2p-1}%
\right] }^{\left[ ii_{1}...i_{2s}...i_{2p-1}\right] }\hat{R}%
_{i_{1}i_{2}}^{j_{1}j_{2}}...\hat{R}%
_{i_{2s-1}i_{2s}}^{j_{2s-1}j_{2s}}K_{i_{2s+1}}^{j_{2s+1}}...K_{i_{2p-1}}^{j_{2p-1}},
\label{PIspij}
\end{equation}%
and $K_{ij}$ is the extrinsic curvature of the $r=const.$ submanifolds. The coefficients $C_{s(p)}$ are%
\begin{equation}
C_{s(p)}=\frac{4^{p-s}}{s!\left[ 2(p-s)-1\right] !!}.  \label{Csp}
\end{equation}

For the particular case of Chern-Simons gravity, the coefficients $\alpha _{p}$ entering in (\ref{Ilovelock}) are fixed to
\begin{equation}
\alpha _{p}=\frac{n!\left[ 2(n-p)\right] !}{2^{p-1}(n-p)!}.
\label{alphap}
\end{equation}%
For this choice, the Lagrangian in (\ref{Ilovelock}) becomes a
Chern-Simons form satisfying $d{\cal L}=F^{n+1}$, with $F\in
SO(2n,2)$.  We shall not, however, make use of the ``gauge theory"
formulation.

For the choice (\ref{alphap}), the momenta can be rewritten in a more compact form, written in terms
of a continuous parameter
$t\in \lbrack 0,1]$,
\begin{eqnarray}
\pi _{j}^{i}
&=&n \sqrt{-h}\int\limits_{0}^{1}dt\,\delta _{\left[
jj_{1}...j_{2n-1}\right] }^{\left[ ii_{1}...i_{2n-1}\right]
}K_{i_{1}}^{j_{1}}\left({1 \over 2}
R_{i_{2}i_{3}}^{j_{2}j_{3}}(h)-t^{2}K_{i_{2}}^{j_{2}}K_{i_{3}}^{j_{3}}+%
\delta _{i_{2}}^{j_{2}}\delta _{i_{3}}^{j_{3}}\right) \times ... \nonumber\\
&&\times \left({1 \over 2} R_{i_{2n-2}i_{2n-1}}^{j_{2n-2}j_{2n-1}}(h)
-t^{2}K_{i_{2n-2}}^{j_{2n-2}}K_{i_{2n-1}}^{j_{2n-1}}+\delta
_{i_{2n-2}}^{j_{2n-2}}\delta _{i_{2n-1}}^{j_{2n-1}}\right).
\label{Momij}
\end{eqnarray}%
where we have used the Gauss-Codazzi relation in the radial foliation%
\begin{equation}
\hat{R}_{kl}^{ij}=R_{kl}^{ij}(h)-K_{k}^{i}K_{l}^{j}+K_{l}^{i}K_{k}^{j}.
\label{GC}
\end{equation}%

For notational simplicity in what follows we shall omit all
indices. For example, the expression (\ref{Momij}) for $\pi^i_{\
j}$ will be written simply as,
\begin{equation}\label{pi}
\pi = n \sqrt{-h} \int_{0}^1 dt\, K \left( {1 \over 2} R(h) - t^2 KK + 1 \right)^{n-1}.
\end{equation}
It is straightforward to reinsert the indices. Note also that since all tensors have the same number of
covariant and contravariant indices, no signs will be lost when manipulating expressions as (\ref{pi}).

We are now ready to compute the on-shell variation appearing in (\ref{1}),
$\delta I_0 = \int \pi^{i}_{\ j}\, h^{jk} \delta h_{ki}
=\int \pi^{i}_{\ j}\, g^{jk} \delta g_{ki} $, for Chern-Simons gravity.  Note that
the extrinsic curvature $K^i_{\ j}$ in the adapted frame (\ref{ds2radial}) takes the simple form
\begin{equation}
K_{\ i}^{ j}=\delta _{i}^{j}-rk_{\ i}^{ j}, \ \ \ \ \ \ \mbox{with} \ \ \ \ \ \   k_{i}^{j}=g^{jl}g_{li}^{\prime}
\label{K=delta+}
\end{equation}%
and the Riemann tensor
\begin{equation}
R_{kl}^{ ij}(h)=r\,R_{kl}^{ ij}(g).
\label{Rh=rRg}
\end{equation}%

Inserting this form for $K$ and $R(h)$ into (\ref{1}), and using (\ref{pi}), we obtain
\begin{equation}\label{En}
\delta I^{(n)}_0=\int_{r=\varepsilon} d^{2n}x\,{n\over r^n}\,\int_0^1 dt\
\sqrt{-g}\,(1-rk)\left({r\over 2}R(g)-t^2(1-rk)^2 + 1 \right)^{n-1} g^{-1}\delta g,
\end{equation}

This formula has a remarkable structure. As an example we display
here the first few values of $n=2,3,4$ corresponding to dimensions
$D=5,7,9$ (keeping only the finite and divergent terms in the
limit $\varepsilon\rightarrow 0$),
\begin{eqnarray}
\delta I^{(2)}_0   &=& \int_{r=\varepsilon} d^{4}x\,
\sqrt{-g}\,\left( k^2+{1\over 2}R k-{1\over 2r}R-{2\over 3r^2}\right)g^{-1}\delta g\nonumber\\
\delta I^{(3)}_0 &=& \int_{r=\varepsilon} d^{6}x\,\sqrt{-g}\,\left( {1 \over 4}k R^2 + k^2 R
+{4 \over 3}k^3-{1\over 4r}R^2-{2 \over 3r^2 }R - {8 \over 15r^3} \right) g^{-1}\delta g\nonumber\\
\delta I^{(4)}_0&=& \int_{r=\varepsilon} d^{8}x\,\sqrt{-g}\,\left(  {3 \over 4 }k^2R^2 + {1 \over 8}kR^3 + 2k^3R
+ 2k^4-{1 \over 8r}R^3-{1\over 2r^2}R^2-{4 \over 5r^3}R - {16\over 35r^4} \right) g^{-1}\delta g  \nonumber\\
\delta I^{(5)}_0&=& ... \label{stru}
\end{eqnarray}
We note that the divergent terms (as $\varepsilon\rightarrow 0)$
only depend on $g_{ij}$ and its associated curvature $R(g)$, but not on $k$. This means that these
terms can be written as total variations. In fact, reinserting the indices,
these terms contain the product of $p$ curvatures which can easily be written as a total variation,
\begin{eqnarray}
\sqrt{-g}\, R^p\, g^{-1}\delta g &=& \sqrt{-g}\delta _{\left[
jj_{1}...j_{2p-1}\right] }^{\left[ ii_{1}...i_{2p-1}\right] }
R_{i_{1}i_{2}}^{j_{1}j_{2}}...R_{i_{2p-2}i_{2p-1}}^{j_{2p-2}j_{2p-1}}
g^{jl}\delta g_{li} \nonumber\\ \label{Rpdeltag}
  &=&2\delta \left(\sqrt{-g}\delta _{\left[
j_{1}...j_{2p}\right] }^{\left[ i_{1}...i_{2p}\right] }
R_{i_{1}i_{2}}^{j_{1}j_{2}}...R_{i_{2p-1}i_{2p}}^{j_{2p-1}j_{2p}}\right)
\label{deltaRp}
\end{eqnarray}
with $p=0\dots n-1$. Note that the second line is valid up to a boundary term.
The counterterm $B$ is thus a local functional of $g_{ij}$ of the Lovelock type in $2n$ dimensions.
We give the explicit form below.

Let us now prove that the structure displayed in (\ref{stru}) is a general property of Chern-Simons
gravities present for all dimensions. We go back to Eq. (\ref{En}). Our aim is to prove that the
divergent terms in this expression do not contain $k$.  To this end, we shall take the derivative
of (\ref{En}) with respect to $k$, and prove that it gives a finite quantity. We compute,
\begin{eqnarray}
\frac{\partial (\delta I_0)}{\partial k} &=&-n\int_{r=\varepsilon} d^{2n}x\,\frac{\sqrt{-g}}{r^{n-1}}%
\int_{0}^{1}dt\left[ \left( \frac{r}{2}R-t^{2}(1-rk)^{2}+1\right)
^{n-1}\right. -  \nonumber \\
&&-\left. 2t^{2}(1-rk)^{2}\left( \frac{r}{2}R-t^{2}(1-rk)^{2}+1\right) ^{n-2}%
\right] g^{-1}dg
\end{eqnarray}%
We see that the integrand is a total derivative respect to $t$
\begin{equation}
\frac{d}{dt}\left[ t\left( \frac{r}{2}R-t^{2}(1-rk)^{2}+1\right) ^{n-1}%
\right].
\end{equation}%
This means that the integral over $t$ can be perform explicitly and we get
\begin{eqnarray}\label{fin}
\frac{\partial (\delta I_0)}{\partial k} &=&-n\int_{r=\varepsilon} d^{2n}x\,
\sqrt{-g}\left( \frac{1}{2}R+2k+rk^{2}\right) ^{n-1}g^{-1}dg
\end{eqnarray}
which is explicitly finite in the limit $\varepsilon\rightarrow 0$.

The piece in $\delta I_0$ that depends on $k$ is thus finite, and can be evaluated at $\varepsilon=0$ directly.
Integrating (\ref{fin}) we find a simple formula for the finite piece
\begin{equation}
\delta I_{fin}=-n\,\int d^{2n}x\sqrt{-g_{(0)}}\int_{0}^{1}dt\, k\left(
{1 \over 2}R_{(0)}+2tk\right) ^{n-1}g_{(0)}^{-1}\delta g_{(0)}
\end{equation}
where $g_{(0)ij} = g_{ij}$ evaluated at $r=0$.
Putting back all indices and varying with respect  to $g_{(0) ij}$, we finally reach
at the general formula for the holographic energy momentum tensor,
\begin{eqnarray}
T^{i}_{j} = {g_{(0) jl} \over 2 \sqrt{-g_{(0)}}}{\delta I_{fin}
\over \delta g_{(0) li}} &=& n\int\limits_{0}^{1}dt\delta _{\left[
jj_{1}...j_{2n-1}\right] }^{\left[ ii_{1}...i_{2n-1}\right]
}k_{i_{1}}^{j_{1}}\left( {1 \over
2}R_{i_{2}i_{3}}^{j_{2}j_{3}}(g)+2tk_{i_{2}}^{j_{2}}\delta
_{i_{3}}^{j_{3}}\right) \times ...  \nonumber\\
&&\times \left( {1 \over
2}R_{i_{2n-2}i_{2n-1}}^{j_{2n-2}j_{2n-1}}(g)+2tk_{i_{2n-2}}^{j_{2n-2}}\delta
_{i_{2n-1}}^{j_{2n-1}}\right).  \label{Tijfinite}
\end{eqnarray}%
It is direct to see that this formula reproduces the finite pieces
in the above variations.  This formula is in full agreement with
the result of \cite{BST}. As shown in that reference, using the
equations of motion, the trace of $T^{ij}$ can be written purely
in terms of $g_{(0)ij}$, and it is equal to the $2n$-dimensional
Euler density
\begin{eqnarray}
T^{i}_{\ i} = {1 \over 2^{n}}\delta_{\left[ j_{1}...j_{2n}\right]
}^{\left[ i_{1}...i_{2n}\right]
}R_{i_{1}i_{2}}^{j_{1}j_{2}}(g_{(0)})...R_{i_{2n-1}i_{2n}}^{j_{2n-1}j_{2n}}(g_{(0)}).
\label{Tii}
\end{eqnarray}%

We end by displaying the explicit formula for $B$. Reinserting all
indices the formula reproducing the divergent pieces in $\delta
I_0$ is,
\begin{eqnarray}
\delta B &=&\frac{n}{2^{n-1}}\frac{\sqrt{-g}}{\varepsilon^{n}}\int%
\limits_{0}^{1}dt\delta _{\left[ jj_{1}...j_{2n-1}\right]
}^{\left[ ii_{1}...i_{2n-1}\right] }\delta _{i_{1}}^{j_{1}}\left(
\varepsilon R_{i_{2}i_{3}}^{j_{2}j_{3}}+2\left( 1-t^{2}\right) \delta
_{i_{2}}^{j_{2}}\delta _{i_{3}}^{j_{3}}\right) \times ...   \nonumber\\
&&\times \left( \varepsilon
R_{i_{2n-2}i_{2n-1}}^{j_{2n-2}j_{2n-1}}+2\left( 1-t^{2}\right)
\delta _{i_{2n-2}}^{j_{2n-2}}\delta _{i_{2n-1}}^{j_{2n-1}}\right)
g^{jl}\delta g_{li}. \label{deltaBct}
\end{eqnarray}%
Now, (\ref{deltaBct}) is exactly the variation of
an action of the Lovelock type. In fact, this can be integrated to yield,
\begin{eqnarray}
B &=&2n(n-1)!  \sqrt{-g}\sum\limits_{p=0}^{n-1}
{2^{n-2p-1}(2(n-p)-1)!! \over \varepsilon^{n-p}} \delta_{\left[
j_{1}...j_{2p}\right] }^{\left[ i_{1}...i_{2p}\right]
}R_{i_{1}i_{2}}^{j_{1}j_{2}}...R_{i_{2p-1}i_{2p}}^{j_{2p-1}j_{2p}}
\label{Bct1}
\end{eqnarray}%
which has exactly the form (\ref{Ilovelock}).
We finally note that
this counterterm action can be expressed in terms of the metric
$h_{ij} = g_{ij}/\varepsilon$, and all dependence on the cutoff parameter
$\varepsilon$ disappears. This is presumably related to the character of the
anomaly which is of type $A$, with no contributions from the Weyl
tensor.

To summarize, we have shown that Chern-Simons gravity in $D=2n+1$-dimensional
 spacetimes has special features which allow the computation of the holographic
 energy momentum tensor explicitly for all $n$. We have also isolated the general
form of the counterterm that renders the action finite, and show
that it has the form of a Lovelock action in $2n$ dimensions.

~

\noindent {\bf Acknowledgments}

MB was partially supported by Fondecyt grants \# 1020832 and \#
7020832, RO by Fondecyt grant \#3030029 and ST by GIF - the
German-Israeli Foundation for Scientific Research.


\end{document}